\documentclass[final]{aipproc}
  
\layoutstyle{8x11single}

\usepackage{bm}
\usepackage{combelow}
\usepackage{slashed}
\usepackage{tensor}
\usepackage{braket}
\usepackage{amsmath}
\usepackage{amssymb}

\def\Et{{\widetilde{E}}}
\def\comm#1#2{\mathinner{\left[{#1}, {#2}\right]}}
\def\acomm#1#2{\mathinner{\left\{{#1}, {#2}\right\}}}

\begin{document}

\title{Dirac fermions on an anti-de Sitter background}
\classification{04.62.+v, 11.10.Wx}
\keywords      {Quantum field theory on curved spaces, Dirac fermions, anti-de Sitter
space-time, bi-spinor of parallel transport, Hadamard renormalization, rigidly rotating thermal states.}

\author{Victor E. Ambru\cb{s}\footnote{app10vea@sheffield.ac.uk}, Elizabeth Winstanley\footnote{E.Winstanley@sheffield.ac.uk}}{
  address={Consortium for Fundamental Physics, School of Mathematics and Statistics, \\ University of Sheffield,
Hicks Building, Hounsfield Road, Sheffield, S3 7RH, United Kingdom}
}

\begin{abstract}
Using an exact expression for the bi-spinor of parallel transport, we construct the Feynman propagator for
Dirac fermions in the vacuum state on anti-de Sitter space-time. We compute the vacuum expectation
value of the stress-energy tensor by removing coincidence-limit divergences using the
Hadamard method. We then use the vacuum Feynman propagator to compute thermal expectation
values at finite temperature. We end with a discussion of rigidly rotating
thermal states.
\end{abstract}

\maketitle

\section{Introduction}\label{sec:intro}

Quantum field theory (QFT) on curved space-time is a semi-classical approximation to quantum gravity in which
quantum fields evolve on a fixed background described by a classical metric.
An object of fundamental importance in QFT on curved space-time is the renormalized expectation value of
the stress-energy
tensor (SET) $\braket{T_{\mu \nu }}_{\rm{ren}}$.
This governs the back-reaction of the quantum field on the space-time geometry via the semi-classical Einstein equations
\begin{equation}
G_{\mu \nu } = 8\pi \braket{T_{\mu \nu }}_{\rm{ren}},
\end{equation}
(we use natural units in which $G=c=\hbar=k_{B}=1$).
The expectation value of the SET and other physical observables are calculating using the Feynman propagator.
This can be found either by construction, using the time-ordered product
of the field operator, or by directly solving the inhomogeneous field equations, using appropriate boundary
conditions.

QFT on curved space-time is considerably more complicated than quantum field theory on Minkowski space-time.
On Minkowski space-time, there is a natural definition of a global vacuum state as seen by an inertial observer.
Defining a vacuum state on a general curved space-time is a subtle procedure and there may be more than one natural choice of vacuum state.
The choice of vacuum state affects the boundary conditions used to construct the appropriate propagator.

In this paper, we focus on the maximally symmetric anti-de Sitter (adS) space. 
According to the adS/CFT (conformal field theory) correspondence (see \cite{art:aharony} for a review),
quantum gravity in the bulk of adS is equivalent to a
CFT which lives on its time-like boundary.
This motivates our study of 
QFT on adS.
We focus on fermion fields as these describe all known matter particles but have received
less attention in the literature than quantum scalar fields on curved spaces.

The maximal symmetry of adS enables us to write the Feynman propagator
in a relatively simple closed form using the bi-spinor of parallel transport \cite{art:muck,art:ambrusksm}.
Considering the global adS vacuum, we use the Hadamard method \cite{art:najmi_ottewill,art:hack}
to calculate the vacuum expectation value (v.e.v.)
of the SET.
We also find the thermal
expectation values (t.e.v.s) of the fermion condensate (FC) and SET.
In addition to the results for the massless Dirac field presented in
Ref.~\cite{art:ambrusksm}, expressions
for the t.e.v.s for arbitrary mass are included.

We next consider the construction of thermal states as seen by
an observer rotating with a constant angular velocity $\bm{\Omega}$ about a fixed axis.
The construction of the vacuum state proceeds in analogy to that for a rotating observer on Minkowski space \cite{art:ambrusrot,art:ambrusmg13}.
When the angular velocity
is smaller than the inverse radius of curvature $\omega$ of adS, the rotating vacuum and its corresponding Feynman
propagator coincide with those for the global non-rotating adS vacuum. In this case, relatively simple expressions
for the t.e.v.s of the FC, SET and neutrino charge current (CC) can be obtained. If $\Omega > \omega$, the rotating and adS
vacua no longer coincide. Since the maximal symmetry of adS is broken by the presence of a prefered axis,
the Feynman propagator must be constructed using a mode sum. In this case, we only give numerical results here
and leave full details to be presented in a dedicated paper \cite{art:ambrusadsrot}.

\section{Geometry of adS}\label{sec:ads-g}

Anti-de Sitter space-time (adS) is a vacuum solution of the Einstein equations
with a negative cosmological constant $\Lambda = -3\omega^2$, where $\omega$ is the inverse radius
of curvature of adS. We use coordinates such that the line element reads as:
\begin{equation}\label{eq:ds2}
 ds^2 = \frac{1}{\cos^2\omega r}
 \left[-dt^2 + dr^2 + \frac{\sin^2\omega r}{\omega^2} \left(d\theta^2 + \sin^2\theta d\varphi^2\right)\right].
\end{equation}
The time coordinate $t$ runs from $-\infty$ to $\infty$, thereby giving the covering space of adS. The radial
coordinate $r$ runs from $0$ to the space-like boundary at $\pi/2\omega$, while $\theta$ and $\varphi$
are the usual elevation and azimuthal angular coordinates.
We introduce the following natural tetrad \cite{art:cota}:
\begin{subequations}\label{eq:tetrad}
\begin{align}
 \omega^{\hat{t}} =& \frac{dt}{\cos\omega r}, &
 \omega^{\hat{i}} =& \frac{dx^j}{\cos\omega r}\left[\frac{\sin\omega r}{\omega r}
 \left(\delta_{ij} - \frac{x^ix^j}{r^2}\right) + \frac{x^ix^j}{r^2}\right],\\
 e_{\hat{t}} =& \cos\omega r\, \partial_t, &
 e_{\hat{i}} =& \cos\omega r\left[\frac{\omega r}{\sin\omega r}
 \left(\delta_{ij} - \frac{x^ix^j}{r^2}\right) + \frac{x^ix^j}{r^2}\right] \partial_j,
\end{align}
\end{subequations}
such that $\eta_{\hat{\alpha}\hat{\beta}} \omega^{\hat{\alpha}}_\mu \omega^{\hat{\beta}}_\nu = g_{\mu\nu}$, where
$\eta_{\hat{\alpha}\hat{\beta}} = \rm{diag}(-1,1,1,1)$ is the Minkowski metric.

The geodetic interval $s(x,x')$ represents the distance between two points with coordinates
$x$ and $x'$ along the geodesic connecting these points. On adS, it is possible to calculate $s$ explicitly \cite{art:allen_jacobson}:
\begin{equation}
 \cos\omega s = \frac{\cos\omega \Delta t}{\cos\omega r\cos\omega r'} - \tan\omega r \tan\omega r' \cos\gamma,
\end{equation}
where $\gamma$ is the angle between $\bm{x}$ and $\bm{x'}$
($\cos\gamma =\cos\theta \cos\theta' + \sin\theta \sin\theta' \cos\Delta \varphi$).
Here, $s$ is real if the geodesic connecting $x$ and $x'$ is time-like and imaginary if it is space-like.
Furthermore, $n_\mu(x,x') = \nabla_\mu s(x,x')$ and $n_{\mu'}(x,x') = \nabla_{\mu'} s(x,x')$ are
time-like vectors tangent to this geodesic at $x$ and $x'$, respectively, obeying
$n_\mu n^\mu = n_{\mu'} n^{\mu'} = -1$.

In general relativity, fields at different space-time points $x$ and $x'$ must first be parallel transported to
the same point before they can be compared. To this end, geodesic theory defines the bi-vector
$g\indices{^\mu_{\nu'}}(x,x')$ and bi-spinor $\Lambda(x,x')$ of parallel transport, which perform the
parallel transport of tensors and spinors, respectively, from point $x'$ to point $x$ along the connecting geodesic.
The bi-vector and bi-spinor of parallel transport therefore satisfy the parallel transport equations:
\begin{equation}
 n^\mu \nabla_\mu g\indices{^\nu_{\lambda'}} = 0, \qquad
 n^\mu D_\mu \Lambda(x,x') = 0,
\end{equation}
where $D_\mu = \partial_\mu + \Gamma_\mu$ is the covariant derivative for spinors and $\Gamma_\mu$ is the
spin connection. On adS, the bi-spinor of parallel transport satisfies the additional equation \cite{art:muck}:
\begin{equation}\label{eq:dlambda}
 D_\mu \Lambda(x,x') = \frac{\omega}{2}\tan\frac{\omega s}{2} (\gamma_\mu \slashed{n} + n_\mu) \Lambda(x,x'),
\end{equation}
where the covariant gamma matrices $\gamma_\mu = \omega_\mu^{\hat{\alpha}} \gamma_{\hat{\alpha}}$
are written in terms of the Minkowski gamma matrices $\gamma^{\hat{\alpha}}$, which obey canonical anti-commutation relations:
\begin{equation}
 \acomm{\gamma^{\hat{\alpha}}}{\gamma^{\hat{\beta}}} = -2\eta^{\hat{\alpha}\hat{\beta}},
\end{equation}
and $\slashed{n} = \gamma^\mu n_\mu$ is the Feynman slash notation.
Without presenting the algebraic details of its construction, the solution of
Eq.~\eqref{eq:dlambda} is \cite{art:ambrusksm,art:ambrusads}:
\begin{multline}
 \Lambda(x,x') = \frac{\cos(\omega\Delta t/2)}{\cos(\omega s/2)\sqrt{\cos\omega r \cos \omega r'}}
 \left\{\cos\frac{\omega r}{2} \cos\frac{\omega r'}{2} +
 \frac{\mathbf{x}\cdot \hat{\mathbf{\gamma}}}{r} \frac{\mathbf{x'}\cdot \hat{\mathbf{\gamma}}}{r'}
 \sin\frac{\omega r}{2} \sin\frac{\omega r'}{2}\right.\\
 \left.- \gamma^{\hat{t}} \tan\frac{\omega\Delta t}{2} \left(
 \frac{\mathbf{x}\cdot \hat{\mathbf{\gamma}}}{r} \sin\frac{\omega r}{2} \cos\frac{\omega r'}{2} -
 \frac{\mathbf{x'}\cdot \hat{\mathbf{\gamma}}}{r'} \cos\frac{\omega r}{2} \sin\frac{\omega r'}{2}\right)\right\}.
 \label{eq:lambda}
\end{multline}
In the next section, the bi-spinor of parallel transport appears in the Feynman propagator $S_F(x,x')$.
The closed-form expression~\eqref{eq:lambda} is needed in the computation of thermal expectation values later in this paper.

\section{Non-rotating fermions}
To keep the present paper self-contained, this section briefly reviews the results in
Refs.~\cite{art:ambrusksm,art:ambrusads} regarding the geometric construction of the Feynman
propagator \cite{art:muck} using the bi-spinor of parallel transport \eqref{eq:lambda} and its
renormalization using the Hadamard method. In the conclusion of this section, we discuss
thermal states where we present novel results regarding thermal expectation values of fermions
of arbitrary mass.

\subsection{AdS Feynman propagator}
The Feynman propagator $S_F(x,x')$ is a solution of the inhomogeneous Dirac equation:
\begin{equation}\label{eq:sf_def}
 (i\slashed{D} - \mu) S_F(x,x') = (-g)^{-1/2} \delta^4(x,x'),
\end{equation}
where $\mu$ is the mass and $g$ is the determinant of the adS metric \eqref{eq:ds2}.
Following Ref.~\cite{art:muck}, the maximal symmetry of adS can be used to cast the Feynman propagator
in the following form:
\begin{equation}\label{eq:sf_muck}
 S_F(x,x') = \left[\alpha_F(s) + \slashed{n}\, \beta_F(s)\right]\, \Lambda(x,x').
\end{equation}
The functions $\alpha_F$ and $\beta_F$ can be shown to obey the following differential equations:
\begin{gather}\label{eq:ab-eqs}
 \left\{\frac{d^2}{d(\omega s)^2} + 3 \cot\omega s\, \frac{d}{d(\omega s)} + \left[k^2 - \frac{3}{2} \left(
 \frac{3}{2} + \frac{\tan(\omega s/2)}{\sin\omega s}\right)\right]\right\} \alpha_F =
 -\frac{k}{\omega} \frac{\delta^4(x,x')}{\sqrt{-g}},\\
 \beta_F = \frac{i}{k} \left(\frac{d}{d(\omega s)} - \frac{3}{2} \tan \frac{\omega s}{2}\right) \alpha_F,
\end{gather}
where $k = \mu / \omega$.
The Feynman propagator is inherently singular as the points $x$ and $x'$ are brought together.
To investigate this singularity, it is convenient to express the solutions of the above equations as follows:
\begin{subequations}\label{eq:sfshort}
\begin{align}
 \alpha_F =& \frac{\omega^3 k}{16\pi^2} \cos\frac{\omega s}{2}\left\{
 -\frac{1}{\sin^2\frac{\omega s}{2}}
 + 2 (k^2 - 1) \ln \left|\sin\frac{\omega s}{2}\right| {}_2F_1\left(2+k,2-k;2;\sin^2\frac{\omega s}{2}\right)
 \right.\nonumber\\
 &\left. + (k^2 - 1) \sum_{n=0}^\infty \frac{(2+k)_n (2-k)_n}{(2)_n n!} \left(\sin^2\frac{\omega s}{2}\right)^n
 \Psi_n\right\},\label{eq:af}\\
 \beta_F =& \frac{i \omega^3}{16\pi^2}  \sin\frac{\omega s}{2}
 \Bigg\{\frac{1 + k^2 \sin^2(\omega s/2)}{[\sin(\omega s/2)]^4}
 - k^2 (k^2 - 1) \ln \left|\sin\frac{\omega s}{2}\right|
 {}_2F_1\left(2+k,2-k;3;\sin^2\frac{\omega s}{2}\right) \nonumber\\
 &\left. - \frac{k^2 (k^2 - 1)}{2} \sum_{n=0}^\infty \frac{(2+k)_n (2-k)_n}{(3)_n n!}
 \left(\sin^2\frac{\omega s}{2}\right)^n \left(\Psi_n - \frac{1}{2+n}\right)\right\},\label{eq:bf}
\end{align}
where $a_n = \Gamma(a+n)/\Gamma(a)$ is the Pochhammer symbol,
$\Gamma(z) = \int_0^\infty x^{z-1} e^{-x} dx$ is the gamma function,
\begin{equation}
 \Psi_n = \psi(k + n + 2) + \psi(k - n - 1) - \psi(n + 2) - \psi(n + 1)
\end{equation}
\end{subequations}
and $\psi(z) = d\ln [\Gamma(z)] / dz$ is the digamma function. The integration constants have been fixed by
matching with the expression for the Feynman propagator in terms of a mode sum \cite{art:ambrusads}.
For the computation of t.e.v.s, it is more convenient to work with the following representation:
\begin{subequations}\label{eq:sfth}
\begin{align}
 \alpha_F =&
 \frac{\omega^3 \Gamma(2 + k) \Gamma(\frac{1}{2})}
 {16 \pi^2 4^k \Gamma(\frac{1}{2} + k)}
 \frac{\cos(\omega s/2)}{[-\sin^2(\omega s/2)]^{2 + k}}
 \, {}_2F_1\left(1 + k, 2 + k; 1 + 2k; \frac{1}{\sin^2(\omega s/2)} \right),\\
 \beta_F =&
 \frac{i \omega^3 \Gamma(2 + k) \Gamma(\frac{1}{2})}
 {16 \pi^2 4^k \Gamma(\frac{1}{2}+ k)}
 \frac{\sin(\omega s/2)}{[-\sin^2(\omega s/2)]^{2 + k}}
 \, {}_2F_1\left(k, 2 + k; 1 + 2k; \frac{1}{\sin^2(\omega s/2)} \right).
\end{align}
\end{subequations}

\subsection{Hadamard renormalization}
The Hadamard renormalization technique has been extensively studied for scalar fields \cite{art:decanini}.
Since the Dirac equation is
a first order differential equation, it is convenient to introduce the auxiliary bi-spinor $\mathcal{G}_F$,
defined by analogy with flat space-time by \cite{art:najmi_ottewill}:
\begin{equation}
 S_F(x,x') = (i\slashed{D} + \mu) \mathcal{G}_F.
\end{equation}
On adS, $\mathcal{G}_F$ can be written using the bi-spinor of parallel transport \cite{art:ambrusksm,art:ambrusads}:
\begin{equation}
 \mathcal{G}_F(x,x') = \frac{\alpha_F}{\mu} \Lambda(x,x'),
\end{equation}
where $\alpha_F$ is defined in \eqref{eq:sf_muck} and $\Lambda(x,x')$ is given by Eq.~\eqref{eq:lambda}.
According to Hadamard's theorem, the divergent part $\mathcal{G}_H$ of $\mathcal{G}_F$ is state-independent
and has the form \cite{art:najmi_ottewill}:
\begin{equation}
 \mathcal{G}_H(x,x') = \frac{1}{8\pi^2} \left[\frac{u(x,x')}{\sigma} + v(x,x') \ln (M^2 \sigma )\right],
\end{equation}
where $u(x,x')$ and $v(x,x')$ are finite when $x'$ approaches $x$, $\sigma = -s^2/2$ is Synge's world function and
$M$ is an arbitrary mass scale.
The bi-spinors $u$ and $v$ can be found by solving the inhomogeneous Dirac equation (\ref{eq:sf_def}), requiring that
the regularized auxiliary propagator $\mathcal{G}_F^{\rm{reg}} \equiv \mathcal{G}_F - \mathcal{G}_H$ is finite
in the coincidence limit:
\begin{subequations}
\begin{align}
 u(x,x') =& \sqrt{\Delta(x,x')} \Lambda(x,x'),\label{eq:uhad}\\
 v(x,x') =& \frac{\omega^2}{2} (k^2 - 1) \cos\frac{\omega s}{2}
 {}_2F_1\left(2-k,2+k;2;\sin^2\frac{\omega s}{2}\right) \Lambda(x,x'), \label{eq:vhad}
\end{align}
\end{subequations}
where the Van Vleck-Morette determinant $\Delta(x,x') = (\omega s / \sin \omega s)^3$ on adS.
The renormalized Feynman propagator can now be written as:
\begin{equation}
 S_F^{\rm{ren}}(x,x') = (i \slashed{D} + \mu) (\mathcal{G}_F - \mathcal{G}_H).
\end{equation}

\subsection{Vacuum stress-energy tensor}
To compute the vacuum expectation value (v.e.v.) of the SET, the following definition must be used \cite{art:hack}:
\begin{equation}\label{eq:tmunu_sf_hack}
 \braket{T_{\lambda\nu}} = \lim_{x'\rightarrow x} {\rm{tr}}
 \left\{\frac{i}{2} \gamma_{(\lambda} \left[D_{\nu)} - g\indices{^{\kappa'}_{\nu )}} D_{\kappa'}\right] S^{\rm{reg}}_F(x,x') +
 \frac{1}{6} g_{\lambda\nu}
 \left[\frac{i}{2} (\slashed{D} - \slashed{D}') - \mu\right] S^{\rm{reg}}_F(x,x') \right\} \Lambda(x',x).
\end{equation}
The bi-vector and bi-spinor of parallel transport are introduced into the above expression so that the Feynman propagator and its
derivatives are correctly evaluated at $x$ \cite{art:groves}.
The definition \eqref{eq:tmunu_sf_hack} differs from the canonical expression by a term proportional to
$g_{\lambda \nu} \mathcal{L}$, where $\mathcal{L}$ is the Dirac Lagrangian. The Lagrangian of the Dirac
field vanishes when the propagator is a solution of the Dirac equation, but in this case, $S^{\rm{reg}}_F$
does not satisfy the Dirac equation, due to the subtraction of its singular part.
The v.e.v. obtained from \eqref{eq:tmunu_sf_hack} matches perfectly the result
obtained by Camporesi and Higuchi \cite{art:camporesi_higucci} using the zeta-function regularization
method \cite{art:ambrusksm,art:ambrusads}. In particular, the trace of the SET is ($\gamma$ is Euler's constant):
\begin{equation}\label{eq:tvac:had}
 \braket{T}_{\rm{vac}}^{\rm{Had}} = -\frac{\omega^4}{4\pi^2}\left\{
 \frac{11}{60} + k - \frac{7k^2}{6} - k^3 + \frac{3k^4}{2} +
 2k^2(k^2 - 1)\left[\ln \frac{M e^{-\gamma} \sqrt{2}}{\omega} - \psi(k)\right]\right\}.
\end{equation}
Although the conformal character of the massless ($k=0$) Dirac field should imply a vanishing trace of the SET,
the renormalization procedure has introduced the so-called conformal anomaly by shifting this trace to
a finite value.
We emphasize that the omission of the term proportional to $g_{\lambda\nu}$ in \eqref{eq:tmunu_sf_hack}
would increase the value of the conformal anomaly by a factor of three.

\subsection{Thermal expectation values}
Thermal expectation values (t.e.v.s) with respect to a thermal bath at inverse temperature $\beta$
can be calculated relative to the vacuum state using the
difference $\Delta S_F^\beta$ between the thermal Feynman propagator $S_F^\beta$ and the vacuum Feynman propagator $S_F$.
This difference can be written as:
\begin{equation}\label{eq:sfb}
 \Delta S^\beta_F(x,x') = \sum_{j\neq 0} S_F(t + ij\beta, \bm{x}; t', \bm{x'}),
\end{equation}
where the functions $\alpha_F$ and $\beta_F$ in $S_F(x,x')$, introduced in Eq.~\eqref{eq:sf_muck},
are in the form given in Eqs.~\eqref{eq:sfth}.
The t.e.v.s we calculate in this section are the fermion condensate (FC) and the SET, given respectively by:
\begin{equation}\label{eq:tmunub}
 \braket{:\overline{\psi}\psi:}_\beta = - \lim_{x'\rightarrow x} {\rm tr} [\Delta S^\beta_F(x,x')],\qquad
 \braket{:T_{\mu\nu}:}_\beta = \frac{i}{2} \lim_{x'\rightarrow x} {\rm{tr}}
 \left\{\left[\gamma_{(\mu} D_{\nu)} - \gamma_{(\mu'}D_{\nu')}\right]\Delta S^\beta_F(x,x')\right\}.
\end{equation}
It can be shown that $\braket{:T\indices{^\mu_\nu}:}_\beta = {\rm diag}(-\rho, p, p, p)$, where
$\rho$ and $p$ are the density and pressure of the Dirac particles, respectively. The results for the t.e.v.s are:
\begin{subequations}\label{eq:tevsads}
\begin{align}
 \braket{:\overline{\psi}\psi:}_\beta =& -\frac{\omega^3 \Gamma_k}{2\pi^2}
 \sum_{j = 1}^\infty (-1)^j \cosh \frac{\omega j \beta}{2} z_{j}^{2+k} {}_2F_1\left(1 + k, 2 + k; 1+ 2k; -z_{j}\right),\\
 \rho + p =& -\frac{\omega^4 \Gamma_k}{4\pi^2}
 \sum_{j = 1}^\infty (-1)^j \cosh\frac{\omega j \beta}{2}\,
 2(2+k) z_{j}^{2+k} {}_2F_1\left(k, 3 + k; 1+ 2k; -z_{j}\right),\\
 p =& -\frac{\omega^4 \Gamma_k}{4\pi^2}
 \sum_{j = 1}^\infty (-1)^j \cosh\frac{\omega j \beta}{2} z_{j}^{2+k} {}_2F_1\left(k, 2 + k; 1+ 2k; -z_{j}\right),
\end{align}
\end{subequations}
where
\begin{equation}\label{eq:zGammak}
 z_{j} = \frac{\cos^2\omega r}{\sinh^2(\omega j \beta/2)}
 \qquad {\rm{and}} \qquad \Gamma_k =  \frac{\Gamma(2+k)\Gamma(\frac{1}{2})}{4^k \Gamma(\frac{1}{2} + k)}.
\end{equation}
In the massless limit $\mu =0$, $k=\mu /\omega$ vanishes, $\Gamma_k = 1$ and Eqs.~\eqref{eq:tevsads} reduce to
\cite{art:ambrusksm,art:ambrusads}:
\begin{subequations}
\begin{align}
 \left.\braket{:\overline{\psi}\psi:}_\beta\right\rfloor_{\mu = 0} =&
 -\frac{\omega^3}{2\pi^2} (\cos\omega r)^4 \sum_{j = 1}^\infty
 \frac{(-1)^j \cosh(\omega j \beta/2)}{[\sinh^2(\omega j \beta / 2) + \cos^2\omega r]^2},\\
 \left.\rho\right\rfloor_{\mu = 0} =& -\frac{3 \omega^4}{4\pi^2} (\cos\omega r)^4 \sum_{j = 1}^\infty (-1)^j
 \frac{\cosh(j \omega \beta/2)}{[\sinh (j \omega \beta / 2)]^4},
\end{align}
\end{subequations}
and $p = \rho / 3$. It is remarkable that the coordinate dependence of the SET is trivially contained
in the $(\cos \omega r)^4$ prefactor.
For $k=0$, profiles of the energy density $\rho $ can be found in Fig.~\ref{fig:rotns} and \cite{art:ambrusksm}.

\section{Rotating fermions}\label{sec:rot}
\begin{figure}
\includegraphics[width=.45\columnwidth]{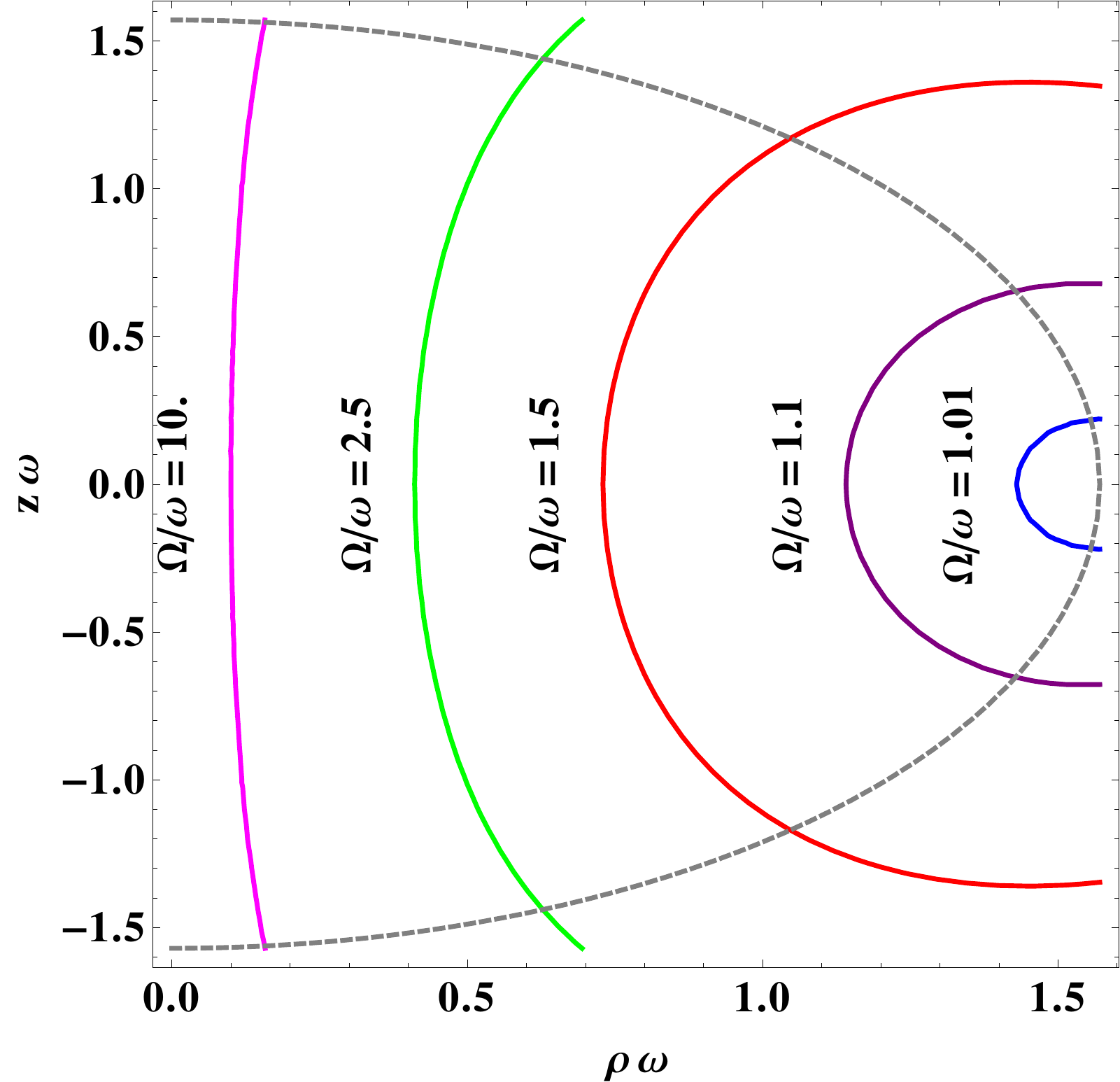}
\caption{Structure of the SOL for various values of $\Omega / \omega$ in a $z$-$\rho$ plot, in units of
$\omega^{-1}$, with the boundary of adS represented by the dotted curve.
The interior of adS is contained within the dotted curve.
The SOL only forms when $\Omega \ge \omega$.
When $\Omega = \omega $ the SOL is situated on the equator of adS at the boundary (i.e.
$r\omega = \theta = \pi / 2$).
As $\Omega$ increases, the SOL approaches the rotation axis and becomes
more cylindrical in shape.}
\label{fig:sols}
\end{figure}

In analogy with the Minkowski case \cite{art:ambrusrot}, let us consider adS as seen by an observer rotating
with a constant angular velocity $\Omega$ about the $z$-axis. The line element in co-rotating coordinates
can be obtained by changing $\varphi \rightarrow \varphi _{R} =\varphi - \Omega t$ in Eq.~\eqref{eq:ds2}:
\begin{equation}\label{eq:ds2-rot}
 ds^2 = \frac{1}{\cos^2\omega r} \left\{-\varepsilon\, dt^2 +
 2 \rho^2 \Omega \left(\frac{\sin\omega r}{\omega r}\right)^2 dt \, d\varphi _{R} + dr^2 +
 \frac{\sin^2\omega r}{\omega^2} d\Omega^2 \right\},
\end{equation}
where $\rho = r \sin\theta$ is the distance from the rotation ($z$) axis and
$\varepsilon = 1 - \rho^2 \Omega^2 \left(\frac{\sin\omega r}{\omega r}\right)^2$.
Under this change of co-ordinates, the $e_{\hat {t}}$ tetrad vector \eqref{eq:tetrad} changes to
\begin{equation}
 e_{\hat{t}} = \cos\omega r\, \left[ \partial_t - \Omega \partial _{\varphi _{R}} \right],
 \label{eq:rottetrad}
\end{equation}
(with the other tetrad vectors unchanged) and the co-frame $ \omega^{\hat{\alpha }}$ also changes accordingly.
The metric \eqref{eq:ds2-rot} reduces to the Minkowski metric in co-rotating coordinates
when $\omega \rightarrow 0$.
The speed of co-rotating particles increases as $\rho$ increases, and equals the speed of light on the
surface where $\varepsilon =0$ (i.e. the speed-of-light surface (SOL)).
It can be seen from the expression for $\varepsilon$ that the SOL only forms if
$\Omega \ge \omega$. Figure~\ref{fig:sols} shows the SOL for different values of
$\Omega / \omega$.

\subsection{Rotating vacuum states}
Geometrically, the metrics \eqref{eq:ds2} and \eqref{eq:ds2-rot} describe the same space-time, using different
coordinates. However, the natural choice of Hamiltonian, $H = i\partial_t$, differs in the two pictures.
The Hamiltonian for the rotating observer $H_{\rm {rot}}$ contains a partial derivative with respect to $t$ holding
$\varphi_{R}=\varphi - \Omega t$ constant whereas the Hamiltonian for a non-rotating observer $H_{\rm {adS}}$ contains a partial derivative with respect to $t$ holding $\varphi $ constant.  Therefore, the two Hamiltonians are related by
$H_{\rm rot} = H_{\rm adS} + i \Omega \partial_\varphi$.

Field modes $U_{\ell }$ have positive frequency if $HU_{\ell } = E_{\ell }U_{\ell }$ with $E_{\ell }>0$.
Since they have different Hamiltonians, the rotating and non-rotating observers have different definitions of positive frequency.
The adS modes $U_{\ell }$ have adS energy $E_{\ell }$, where $H_{\rm {adS}}U_{\ell }=E_{\ell }U_{\ell }$ and
co-rotating modes ${\widetilde{U}}_{\ell }$ have co-rotating energy $\Et_{\ell }$, where $H_{\rm {rot}}{\widetilde{U}}_{\ell } = \Et _{\ell }
{\widetilde {U}}_{\ell }$.
The co-rotating modes $\widetilde{U}_\ell$ can be obtained by performing a coordinate transformation
on the adS modes $U_\ell$, namely:
 $\widetilde{U}_\ell(x) = e^{ i \Omega t M_{z}} U_\ell(x)$,
where $M_{z}$ is the $z$ component of the angular momentum operator.
Choosing the modes $U_{\ell }$ to be simultaneous eigenvectors of $H_{\rm {adS}}$ and $M_{z}$
(on adS, $\comm{H_{\rm {adS}}}{M_{z}} = 0$ \cite{art:cota}), the adS and co-rotating energies are related by
$\Et_\ell = E_\ell - \Omega m_\ell$, where $m_\ell$ is the $z$ component of the
angular momentum of mode $\ell$.
The non-rotating and rotating observers therefore define modes to have positive frequency if $E_{\ell }>0$ and
$\Et_{\ell }>0$ respectively.

Canonical quantization interprets modes with positive frequency as particle modes, while modes with negative frequency are interpreted as
anti-particle modes.
The general solution of the Dirac equation $\psi (x)$ is expanded in terms of field modes and promoted to an operator (here we consider the
co-rotating modes):
\begin{equation}
 \psi(x) = \sum_{\ell} \theta(\Et_\ell) [\widetilde{U}_\ell(x) \widetilde{b}_{\ell} +
 \widetilde{V}_\ell(x) \widetilde{d}^\dagger_{\ell}],
\end{equation}
where the anti-particle modes $\widetilde{V}_\ell$ are the charge conjugates of the particle modes (and therefore have negative frequency):
$\widetilde{V}_\ell = i\gamma^{\hat{\,2}} \widetilde{U}_\ell^*$,
and $\gamma^{\,\hat{2}}$ is the Minkowski $\gamma$ matrix.
The one-particle operators satisfy the canonical anti-commutation relations:
$\acomm{{\widetilde{b}}_\ell}{{\widetilde{b}}^\dagger_{\ell'}} = \acomm{{\widetilde{d}}_\ell}{{\widetilde{d}}^\dagger_{\ell'}} = \delta_{\ell\ell'}$,
$\acomm{{\widetilde{b}}_\ell}{{\widetilde{b}}_{\ell'}} = \acomm{{\widetilde{d}}_\ell}{{\widetilde{d}}_{\ell'}} = 0$.
The vacuum state corresponding to this choice of positive frequency is the state annihilated by all the one-particle annihilation operators
${\widetilde{b}}_{\ell }$, ${\widetilde{d}}_{\ell }$.

The non-rotating and rotating observers have different definitions of positive frequency and, accordingly, will define different vacuum states
by the above procedure.
The non-rotating adS vacuum is analogous to the global Minkowski vacuum,
while the rotating vacuum corresponds to Iyer's quantization \cite{art:iyer} for rotating states in Minkowski space.
Since thermal quantum states are defined relative to a vacuum state (see, for example, \eqref{eq:sfb}), the different rotating and non-rotating
vacua give rise to different t.e.v.s.
A similar effect arises in Minkowski space, where defining rotating thermal states relative to the global Minkowski vacuum gives rise to t.e.v.s containing temperature-independent terms \cite{art:ambrusrot,art:vilenkin} which are not present if the rotating vacuum is used instead.

AdS has a time-like boundary which quantizes the non-rotating mode energies.
As a result of this energy quantization, it can be shown \cite{art:cota,art:ambrusads} that $E_\ell \ge \omega m_\ell$.
Hence, $\Et_\ell \ge 0$ for all values of $\ell$ if $\Omega \le  \omega$ (i.e. when there is no SOL),
in which case the rotating vacuum actually coincides with the adS vacuum.
If $\Omega > \omega $, then the rotating and non-rotating vacua are distinct.

\subsection{Rotating thermal states: $\Omega \le \omega$}
When $\Omega \le \omega$, the rotating and non-rotating vacua coincide. Hence, the vacuum Feynman propagator
$\widetilde{S}_F(x,x')$ can be calculated from the non-rotating propagator $S_F(x,x')$
by applying a coordinate transformation:
\begin{equation}
 \widetilde{S}_F(x,x') = e^{i \Omega t M_z} S_F(x,x') e^{-i\Omega t' M_z},
\end{equation}
where $e^{-i\Omega t' M_z}$ acts from the right on $S_F(x,x')$. The thermal propagator can be constructed
using the method outlined in Eq.~\eqref{eq:sfb}. In addition to the t.e.v.s of the non-rotating case,
there is a non-vanishing current of neutrinos (i.e. particles of negative helicity), which can be calculated using:
\begin{equation}
 \braket{:J^{\hat{\alpha}}_\nu:}_\beta = -\lim_{x'\rightarrow x} {\rm tr} \left[
 \gamma^{\hat{\alpha}} \frac{ 1 - \gamma^5}{2} \Delta \widetilde{S}_F^\beta(x,x')\right].
\end{equation}

\begin{figure}
\begin{tabular}{cc}
\includegraphics[width=.45\columnwidth]{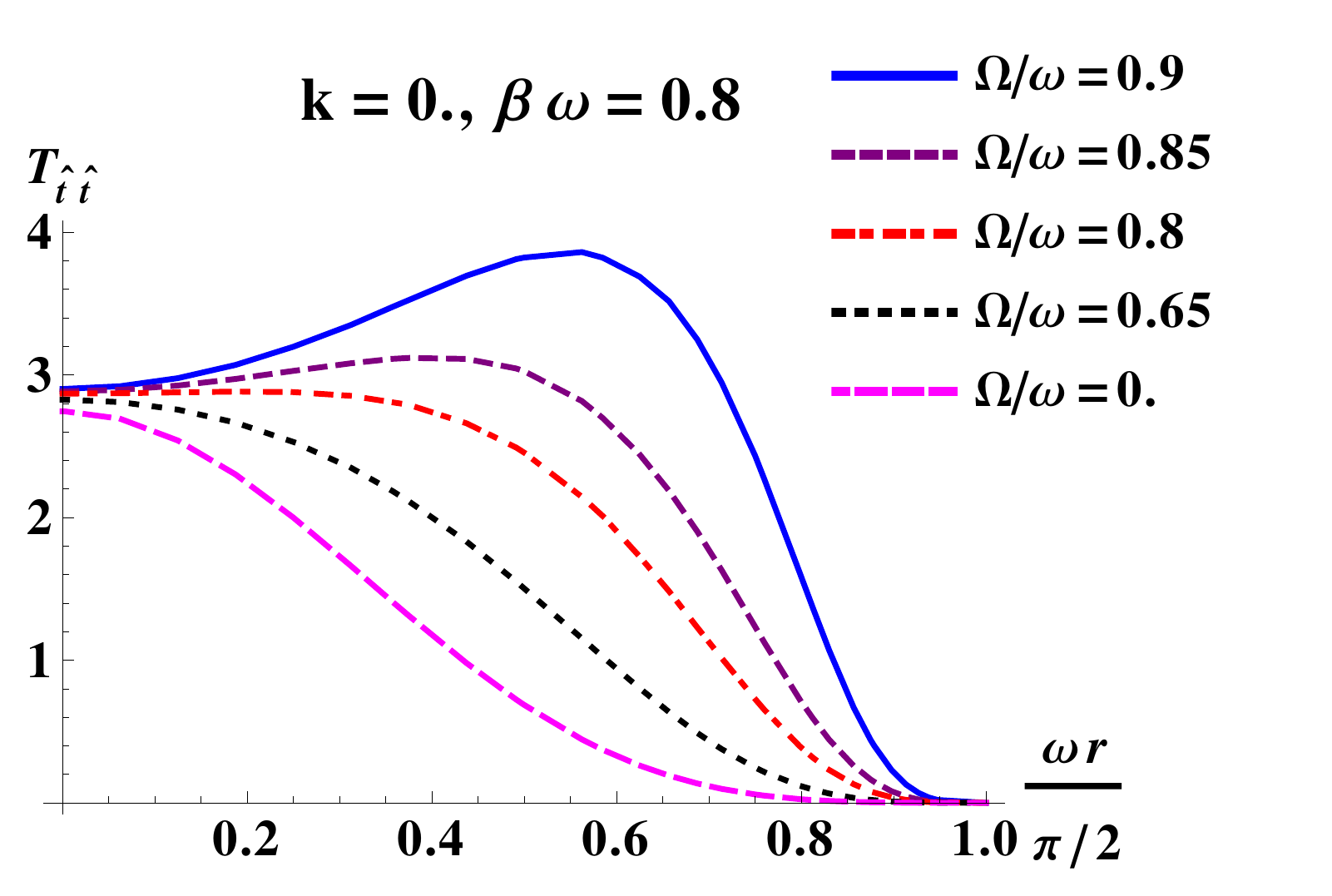} \hspace{.04\columnwidth} &
\includegraphics[width=.45\columnwidth]{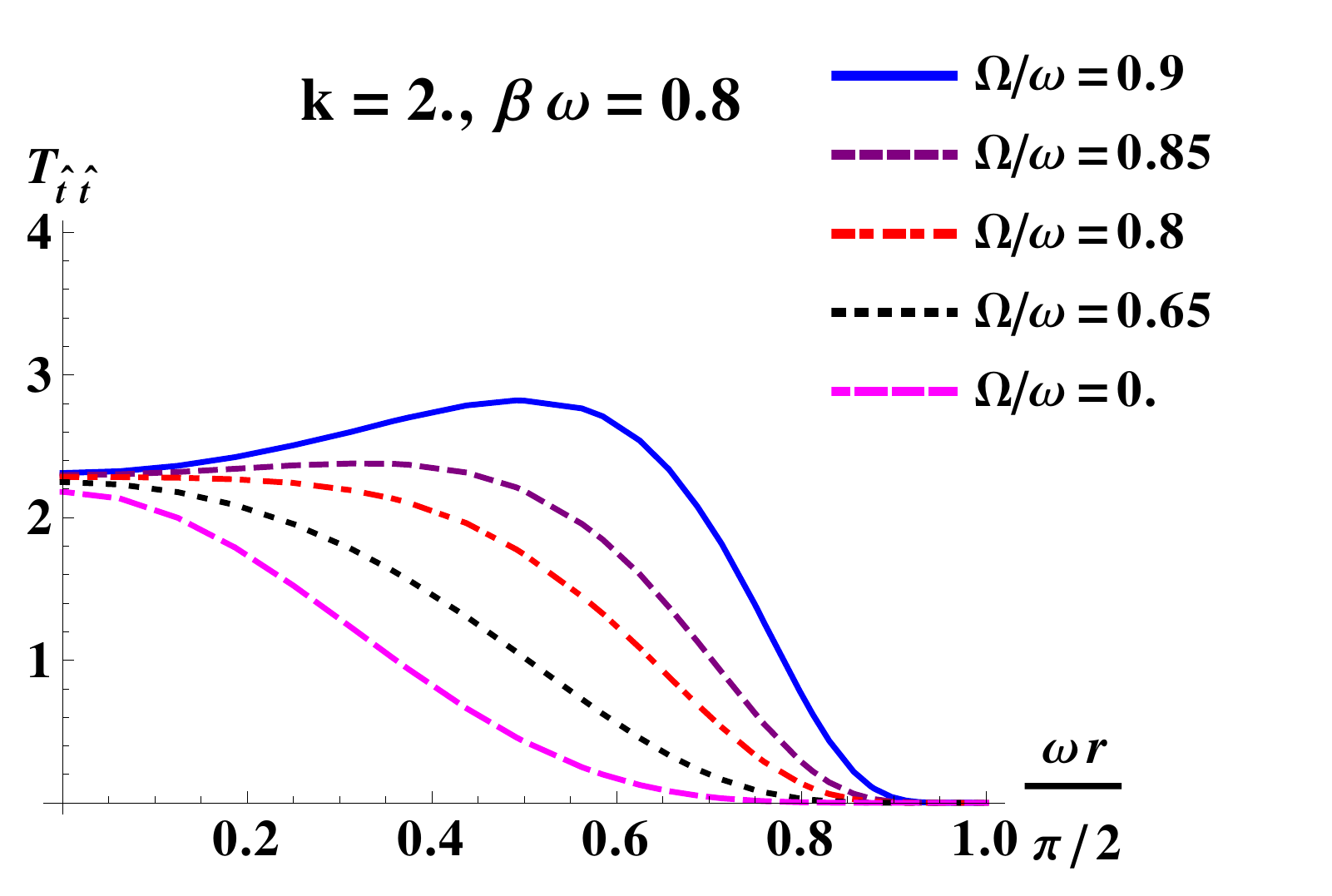}\\
(a) & (b)
\end{tabular}
\caption{Thermal expectation values $\braket{:T_{\hat{t}\hat{t}}:}_\beta$ for fermions of mass
$\mu = 0$ (left) and $\mu = 2\omega$ (right) in the equatorial plane ($\theta = \pi / 2$) for
$\beta \omega = 0.8$ and $\Omega/\omega = 0$ (non-rotating adS), $0.65$, $0.8$, $0.85$ and $0.9$.
The thermal state becomes more energetic as $\Omega$ increases. }
\label{fig:rotns}
\end{figure}

The resulting non-zero t.e.v.s have the following components with respect to the tetrad \eqref{eq:rottetrad}:
\begin{subequations}\label{eq:tevsrot}
\begin{align}
 \braket{:\overline{\psi}\psi:}_\beta =& -\frac{\omega^3 \Gamma_k}{2\pi^2}
 \sum_{j = 1}^\infty (-1)^j \cosh \frac{\omega j \beta}{2} \cosh\frac{\Omega j \beta}{2}
 \zeta_j^{2+k} {}_2F_1\left(1 + k, 2 + k; 1+ 2k; -\zeta_j\right),\\
 \braket{:J^{\hat{r}}_\nu:}_\beta =& \frac{\omega^3 \Gamma_k \cos\theta}{4\pi^2 \cos\omega r}
 \sum_{j = 1}^\infty (-1)^j \sinh\frac{\omega j \beta}{2} \sinh\frac{\Omega j \beta}{2}
 \zeta_j^{2+k} {}_2F_1\left(k, 2 + k; 1+ 2k; -\zeta_j\right),\\
 \braket{:J^{\hat{\theta}}_\nu:}_\beta =& -\frac{\omega^3 \Gamma_k \sin\theta}{4\pi^2}
 \sum_{j = 1}^\infty (-1)^j \sinh\frac{\omega j \beta}{2} \sinh\frac{\Omega j \beta}{2}
 \zeta_j^{2+k} {}_2F_1\left(k, 2 + k; 1+ 2k; -\zeta_j\right),\\
 \braket{:T_{\hat{t}\hat{t}}:}_\beta =& -\frac{\omega^4 \Gamma_k}{4\pi^2\cos^2\omega r}
 \sum_{j = 1}^\infty (-1)^j \cosh\frac{\omega j \beta}{2} \cosh\frac{\Omega j \beta}{2}\nonumber\\
 &\times \left[2(2+k) \sinh^2\frac{\omega j \beta}{2} \zeta_j^{3+k} {}_2F_1\left(k, 3 + k; 1+ 2k; -\zeta_j\right) -
  \cos^2\omega r \,\zeta_j^{2+k} {}_2F_1\left(k, 2 + k; 1+ 2k; -\zeta_j\right)\right],\\
 \braket{:T_{\hat{t}\hat{\varphi}}:}_\beta =& \frac{\omega^4 \Gamma_k \sin\omega r \sin \theta}{8\pi^2 \cos^2\omega r}
 \sum_{j = 1}^\infty (-1)^j \sinh\frac{\omega j \beta}{2} \sinh\frac{\Omega j \beta}{2}\nonumber\\
 &\times 2(2+k) \left(\cosh^2\frac{\omega j \beta}{2} + \cosh^2\frac{\Omega j \beta}{2}\right)
 \zeta_j^{3+k} {}_2F_1\left(k, 3 + k; 1+ 2k; -\zeta_j\right),\\
 \braket{:T_{\hat{r}\hat{r}}:}_\beta =& -\frac{\omega^4 \Gamma_k}{4\pi^2}
 \sum_{j = 1}^\infty (-1)^j \cosh\frac{\omega j \beta}{2} \cosh\frac{\Omega j \beta}{2}
 \zeta_j^{2+k} {}_2F_1\left(k, 2 + k; 1+ 2k; -\zeta_j\right),\\
 \braket{:T_{\hat{\varphi}\hat{\varphi}}:}_\beta =& -\frac{\omega^4 \Gamma_k}{4\pi^2\cos^2 \omega r}
 \sum_{j = 1}^\infty (-1)^j \cosh\frac{\omega j \beta }{2} \cosh\frac{\Omega j \beta }{2} \left[
 \cos^2\omega r\, \zeta_j^{2+k} {}_2F_1\left(k, 2 + k; 1+ 2k; -\zeta_j\right)\right.\nonumber\\
 &\left.+ 2(2+k) \sinh^2\frac{\Omega j \beta}{2} \sin^2 \omega r \sin^2\theta\,
 \zeta_j^{3+k} {}_2F_1\left(k, 3 + k; 1+ 2k; -\zeta_j\right)\right],
\end{align}
\end{subequations}
and $\braket{:T_{\hat{\theta}\hat{\theta}}:}_\beta = \braket{:T_{\hat{r}\hat{r}}:}_\beta$. In \eqref{eq:tevsrot},
\begin{equation}
 \zeta_j = \frac{\cos^2\omega r}{\sinh^2(\omega j \beta / 2) - \sin^2\omega r \sin^2\theta \sinh^2 (\Omega j \beta / 2)},
\end{equation}
and $\Gamma_k$ is given in Eq.~\eqref{eq:zGammak}.
When $\Omega = 0$, $\zeta_j$ reduces to $z_{j}$ defined in Eq.~\eqref{eq:zGammak} and Eqs.~\eqref{eq:tevsrot}
reduce to Eqs.~\eqref{eq:tevsads}.
Equations~\eqref{eq:tevsrot} are regular for any combination of the parameters
(i.e. $k$, $\beta$, $\Omega$, $\omega$, $r$ and $\theta$), as long as $\Omega < \omega$.
Figure~\ref{fig:rotns} shows
$\braket{:T_{\hat{t}\hat{t}}:}_\beta$ at fixed $\beta$ for various values of $\Omega < \omega$, for
both massless and massive fermions. It can be seen that the thermal state becomes more energetic as $\Omega$
increases.

\begin{figure}
\begin{tabular}{cc}
\includegraphics[width=.45\columnwidth]{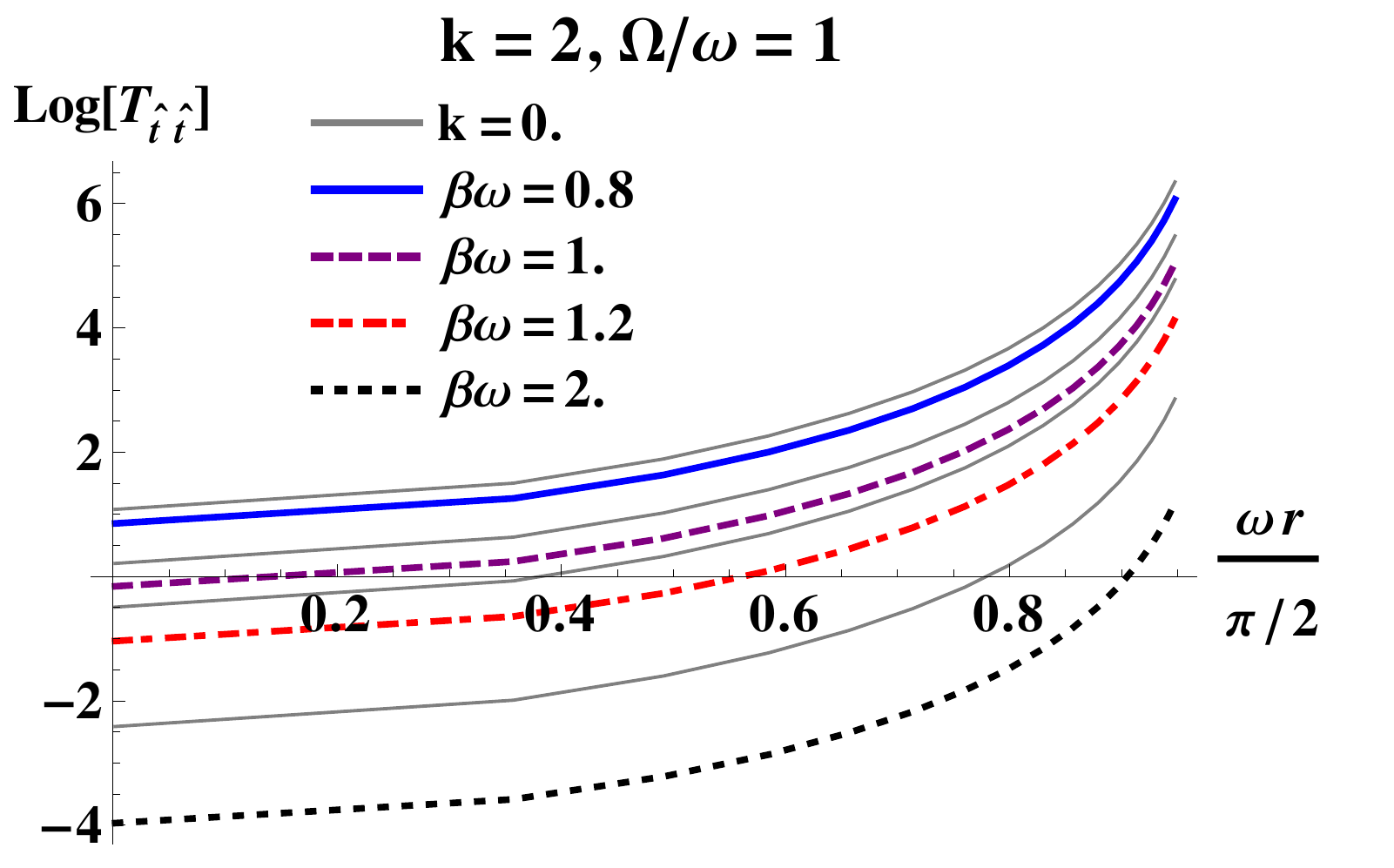} \hspace{.04\columnwidth} &
\includegraphics[width=.45\columnwidth]{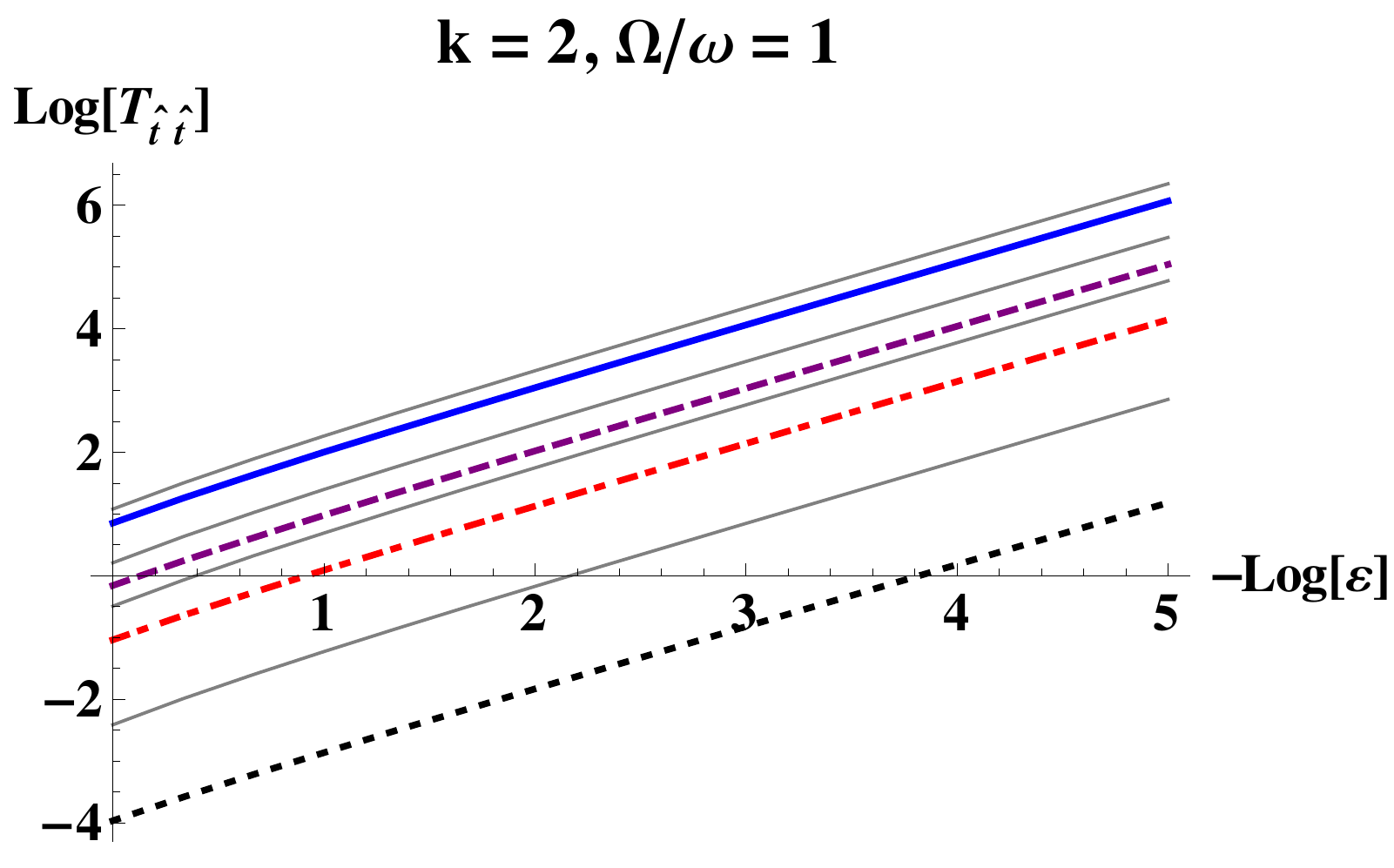}\\
(a) & (b)
\end{tabular}
\caption{Comparison of thermal expectation values $\braket{:T_{\hat{t}\hat{t}}:}_\beta$ of massless (thin lines)
and massive ($\mu = 2\omega$) fermions in the equatorial plane ($\theta = \pi / 2$) for $\Omega/\omega = 1.0$ and
$\beta \omega = 0.8$, $1$, $1.2$ and $2$. Plot (a) shows $\ln \braket{:T_{\hat{t}\hat{t}}:}_\beta$
against the distance from the rotation axis,
while plot (b) shows a logarithmic plot against $\ln \varepsilon^{-1}$, where $\varepsilon = 1 - \sin^2\omega r $.
It can be seen that the divergence is of order $\varepsilon^{-1}$ irrespective of the value of the mass.}
\label{fig:roto1}
\end{figure}

In the special case $\Omega = \omega$,  the tetrad components $\braket{:T_{\hat{t}\hat{t}}:}$, $\braket{:T_{\hat{t}\hat{\varphi}}:}$ and
$\braket{:T_{\hat{\varphi}\hat{\varphi}}:}$ diverge as the adS equator is approached
(i.e. as $\theta \rightarrow \pi/2$ and $\omega r \rightarrow \pi / 2$), due to the $\cos^2\omega r$ factor
in the denominator inside the sum over $j$ (see Fig.~\ref{fig:roto1}).
Equations~\eqref{eq:tevsrot} simplify considerably when $k = 0$ and $\Omega = \omega$:
\begin{subequations}\label{eq:tevsrotk0}
\begin{align}
 \braket{:\overline{\psi}\psi:}_\beta\rfloor_{\substack{\Omega = \omega\\k = 0}} =&
 -\frac{\omega^3(\cos \omega r)^4}{2\pi^2\varepsilon^2}
 \sum_{j = 1}^\infty \frac{(-1)^j \cosh^2(\omega j \beta / 2)}{\displaystyle \left(\cosh^2\frac{\omega j \beta}{2} -
 \frac{\sin^2 \omega r \cos^2\theta}{\cos^2\omega r + \sin^2\omega r \sin^2\theta}\right)^2},\\
 \braket{:J^{\hat{r}}_\nu:}_\beta\rfloor_{\substack{\Omega = \omega\\k = 0}} =&
 -\frac{\omega^3 (\cos\omega r)^3 \cos\theta}{4\pi^2 \varepsilon^2} S_2(\omega \beta),\\
 \braket{:J^{\hat{\theta}}_\nu:}_\beta\rfloor_{\substack{\Omega = \omega\\k = 0}} =&
 \frac{\omega^3 (\cos\omega r)^4 \sin\theta}{4\pi^2 \varepsilon^2} S_2(\omega \beta),\\
 \braket{:T_{\hat{t}\hat{t}}:}_\beta\rfloor_{\substack{\Omega = \omega\\k = 0}} =&
 \frac{\omega^4 (\cos\omega r)^4 (4 - \varepsilon)}{4\pi^2 \varepsilon^3}
 [S_4(\omega\beta) + S_2(\omega\beta)],\\
 \braket{:T_{\hat{t}\hat{\varphi}}:}_\beta\rfloor_{\substack{\Omega = \omega\\k = 0}} =&
 -\frac{\omega^4 (\cos\omega r)^4 \sin\omega r \sin\theta}{\pi^2 \varepsilon^3}
 [S_4(\omega\beta) + S_2(\omega\beta)],\\
 \braket{:T_{\hat{r}\hat{r}}:}_\beta\rfloor_{\substack{\Omega = \omega\\k = 0}} =&
 \frac{\omega^4 (\cos\omega r)^4}{4\pi^2 \varepsilon^2} [S_4(\omega\beta) + S_2(\omega\beta)],\\
 \braket{:T_{\hat{\varphi}\hat{\varphi}}:}_\beta\rfloor_{\substack{\Omega = \omega\\k = 0}} =&
 \frac{\omega^4 (\cos\omega r)^4 (4 - 3\varepsilon)}{4\pi^2 \varepsilon^3}
 [S_4(\omega\beta) + S_2(\omega\beta)],
\end{align}
\end{subequations}
where $S_\ell(x) = -\sum_{j = 1}^\infty (-1)^j/[\sinh(jx/ 2)]^\ell$.
The series $S_{2}(x)$ arising in the tetrad components of the neutrino charge current can be written
using the Q-Pochhammer symbol $(a;q)_k = \prod_{n = 0}^k (1 - aq^n)$ \cite{NIST}:
\begin{equation}
 S_2(x) = -4 \frac{d}{dx} \ln(-1, e^{-x})_\infty.
\end{equation}
It is remarkable that, apart from the fermion condensate, the coordinate dependance for all the t.e.v.s
in Eqs.~\eqref{eq:tevsrotk0} is contained in trigonometric prefactors.

\subsection{Rotating thermal states: $\Omega > \omega$}
We have seen that some t.e.v.s diverge as the inverse distance to the adS equator when $\Omega = \omega$.
If $\Omega$ is further increased, all t.e.v.s become divergent as inverse powers of the distance $\varepsilon$
to the SOL. Since the vacuum state changes  if $\Omega > \omega$, the Feynman propagator method used up until now is no
longer applicable. In this case, the t.e.v.s have to be calculated using mode sums \cite{art:ambrusrot}.

\begin{figure}[b]
\begin{tabular}{cc}
\includegraphics[width=.45\columnwidth]{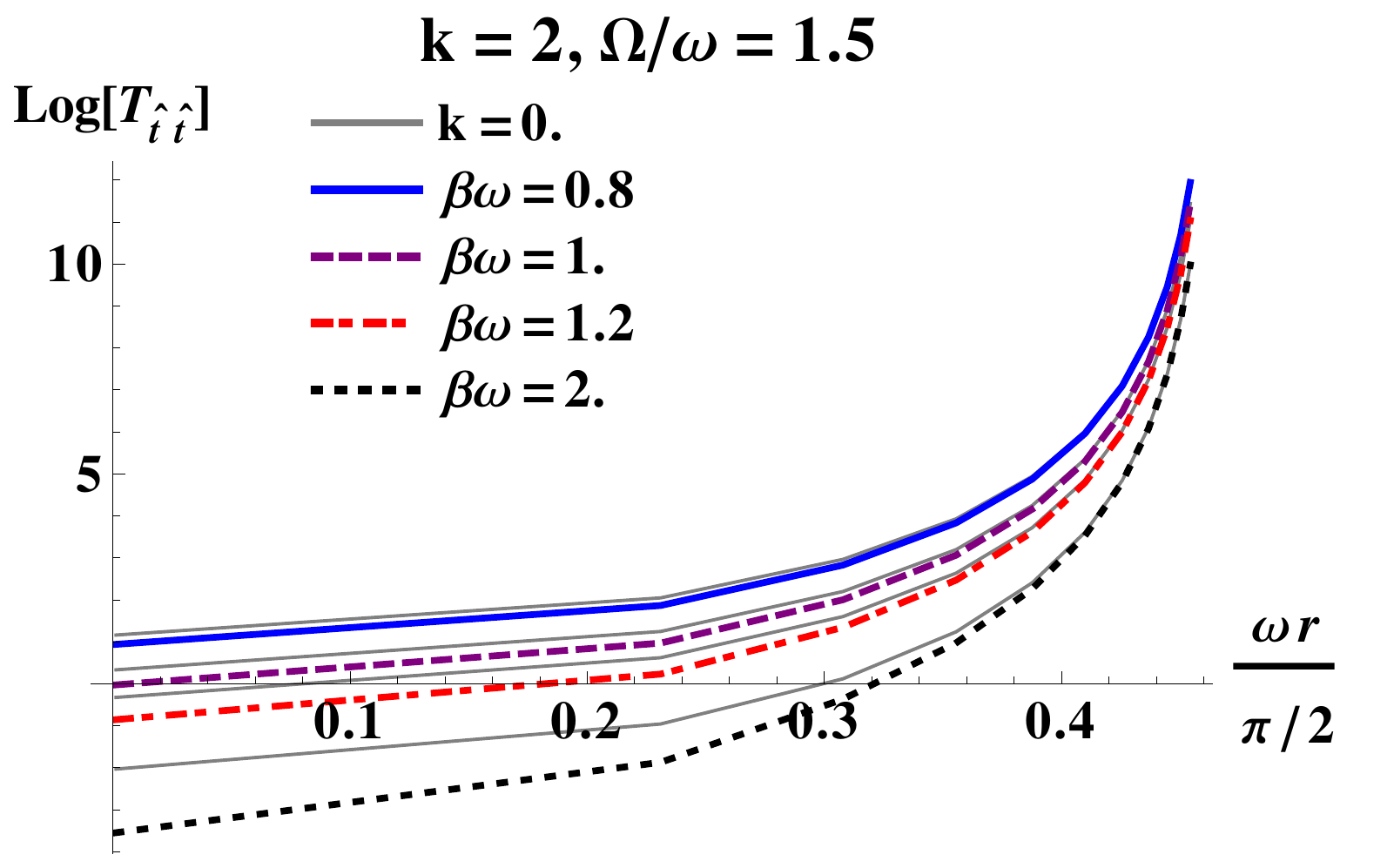} \hspace{.04\columnwidth}&
\hspace{.04\columnwidth}\includegraphics[width=.45\columnwidth]{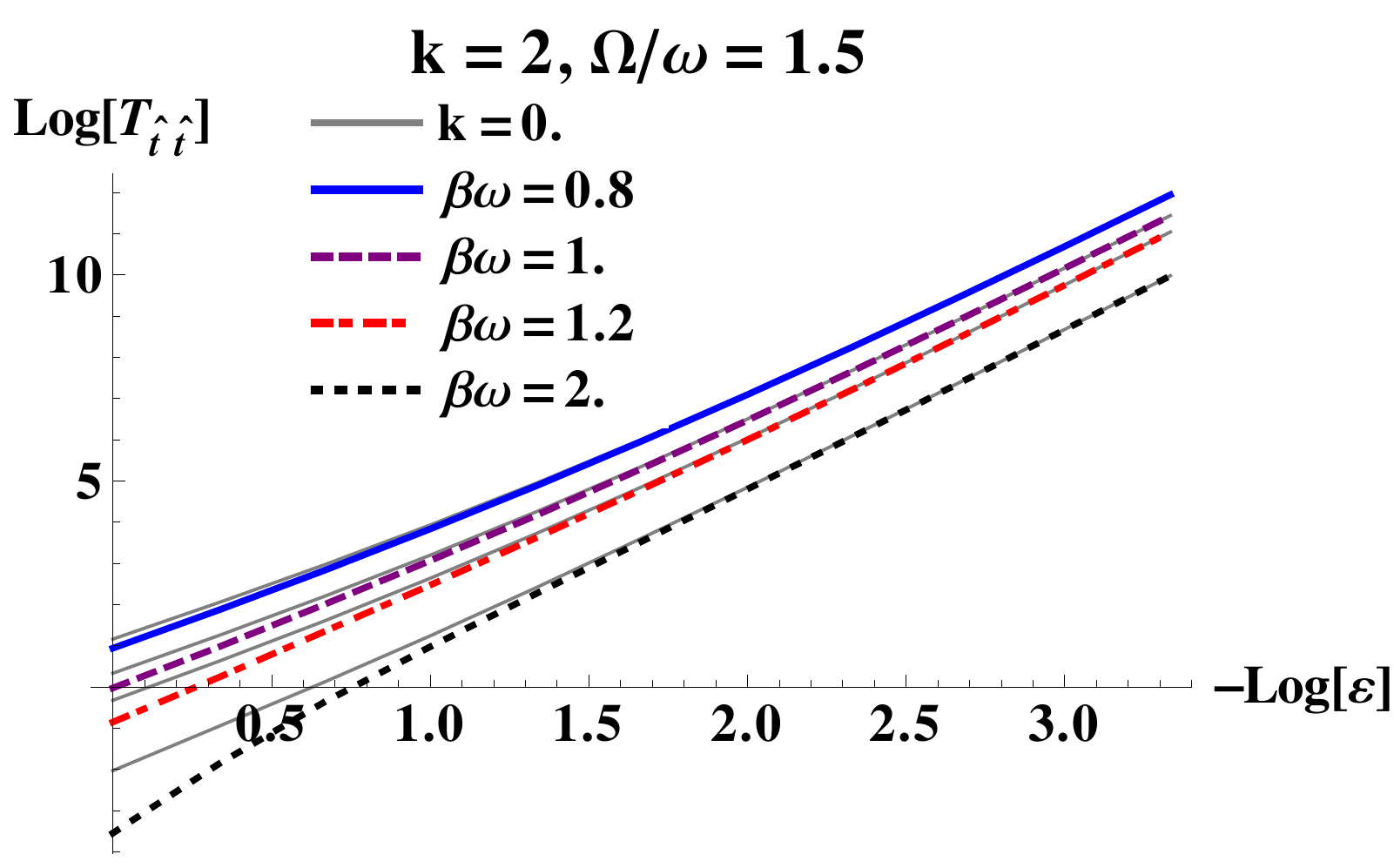}\\
(a) & (b)
\end{tabular}
\caption{Comparison of thermal expectation values $\braket{:T_{\hat{t}\hat{t}}:}_\beta$ of massless (thin lines)
and massive ($\mu = 2\omega$) fermions in the equatorial plane ($\theta = \pi / 2$) for $\Omega/\omega = 1.5$ and
$\beta \omega = 0.8$, $1$, $1.2$ and $2$. Plot (a) shows $\ln \braket{:T_{\hat{t}\hat{t}}:}_\beta$
against the distance from the rotation axis,
while plot (b) shows a logarithmic plot against $\ln \varepsilon^{-1}$.}
\label{fig:rots}
\end{figure}

Figure~\ref{fig:rots} shows the energy density as the SOL is approached, confirming that
$\braket{:T_{\hat{t}\hat{t}}:}_\beta$ diverges as an inverse power of the distance to the SOL.
While further analytic investigations are required to confirm the exact order of this divergence,
numerical results show that its order is higher than in the case $\Omega = \omega$.
The leading order divergence seems to be the same regardless of the mass of the fermions.

\pagebreak

\section{Conclusions}\label{sec:conc}
In this paper, we considered quantum states of fermions of arbitrary mass in anti-de Sitter space (adS)
as seen by an observer rotating with a constant angular velocity $\Omega$.
In the non-rotating case ($\Omega = 0$), the geometric properties of adS were used to construct
the Feynman propagator in closed form, from which vacuum expectation values for the fermion
condensate (FC) and stress-energy tensor (SET) were calculated using Hadamard renormalization.
Thermal expectation values were calculated using a closed form expression for the bi-spinor
of parallel transport.

In the case when $0<\Omega \le \omega$, we found that the rotating and non-rotating vacua were the same.
Thus, the non-rotating vacuum Feynman propagator was used to derive expressions for the thermal expectation values (t.e.v.s) of
the FC, neutrino charge current (CC) and SET. If $\Omega < \omega $, all t.e.v.s stay finite throughout adS.
When $\Omega > \omega$, mode sums were required to
evaluate t.e.v.s.
For these values of $\Omega $ a speed of light surface (SOL) forms and t.e.v.s diverge as inverse powers of the distance to this SOL.
If $\Omega = \omega$, the SOL collapses down to a curve on the equator of adS. In this case, the FC, CC, $T_{\hat{r}\hat{r}}$
and $T_{\hat{\theta}\hat{\theta}}$ stay constant throughout the equatorial plane while $T_{\hat{t}\hat{t}}$,
$T_{\hat{t}\hat{\varphi}}$ and $T_{\hat{\varphi}\hat{\varphi}}$ diverge as
$1/\cos^2\omega r$ as $\omega r \rightarrow \pi/2$.

\begin{theacknowledgments}
This work is supported by the Lancaster-Manchester-Sheffield Consortium for
Fundamental Physics under STFC grant ST/J000418/1,
the School of Mathematics and Statistics at the University of Sheffield
and the European Cooperation in Science and Technology (COST) action MP0905
``Black Holes in a Violent Universe''.
\end{theacknowledgments}

\bibliographystyle{aipproc}

\begin{thebibliography}{99}

\bibitem{art:aharony}
O.~Aharony, S.~S.~Gubser, J.~M.~Maldacena, H.~Ooguri, and Y.~Oz,
Phys.~Rept.~\textbf {323}, 183--386 (2000).

\bibitem{art:muck}
W.~M\"uck, J.~Phys.~\textbf{A 33}, 3021--3026 (2000).

\bibitem{art:ambrusksm}
V.~E.~Ambru\cb{s} and E.~Winstanley, {\emph {Proceedings of the first Karl-Schwarzschild Meeting KSM2013}},
 arXiv:1310.7429 [gr-qc] (2013).

\bibitem{art:najmi_ottewill}
A.-H.~Najmi and A.~C.~Ottewill, Phys.~Rev.~\textbf{D 30}, 2573--2578 (1984).

\bibitem{art:hack}
C.~Dappiaggi, T.-P.~Hack, and N.~Pinamonti, Rev.~Math.~Phys.~\textbf{21}, 1241--1312 (2009).

\bibitem{art:ambrusrot}
V.~E.~Ambru\cb{s} and E.~Winstanley, preprint
arXiv:1401.6388 [hep-th] (2014).

\bibitem{art:ambrusmg13}
V.~E.~Ambru\cb{s} and E.~Winstanley,
{\emph {Proceedings of the thirteenth Marcel Grossman meeting MG13}}, arXiv:1302.3791 [gr-qc] (2013).

\bibitem{art:ambrusadsrot}
V.~E.~Ambru\cb{s} and E.~Winstanley, \emph{Rotating fermions on adS}, paper in preparation.

\bibitem{art:cota}
I.~Cot\u{a}escu, Rom.~J.~Phys.~\textbf{52}, 895--940 (2007).

\bibitem{art:allen_jacobson}
B.~Allen and T.~Jacobson, Commun.~Math.~Phys.~\textbf{103}, 669--692 (1986).

\bibitem{art:ambrusads}
V.~E.~Ambru\cb{s} and E.~Winstanley, \emph{Fermions on adS}, paper in preparation.

\bibitem{art:decanini}
Y.~Decanini and A.~Folacci, Phys.~Rev.~\textbf {D 78}, 044025 (2008).

\bibitem{art:groves}
P.~B.~Groves, P.~R.~Anderson, and E.~D.~Carlson, Phys.~Rev.~\textbf{D 66}, 124017 (2002).

\bibitem{art:camporesi_higucci}
R.~Camporesi and A.~Higuchi, Phys.~Rev.~\textbf{D 45}, 3591--3603 (1992).

\bibitem{art:iyer}
B.~R.~Iyer, Phys.~Rev.~\textbf{D 26}, 1900--1905 (1982).

\bibitem{art:vilenkin}
A.~Vilenkin, Phys.~Rev.~\textbf{D 21}, 2260--2269 (1980).

\bibitem{NIST}
F.~W.~J.~Olver, D.~W.~Lozier, R.~F.~Boisvert, and C.~W.~Clark, {\emph {NIST Handbook of Mathematical Functions}},
(Cambridge University Press, Cambridge, 2010), p.~436.

\end{thebibliography}

\end{document}